\magnification =\magstep 1
\baselineskip 15pt
\rightline {HEP-PH/9509261}
$$ $$
$$ $$
\centerline{\bf  LEPTON  OSCILLATIONS}
$$
$$
\centerline{Y. N. Srivastava, A. Widom and 
E. Sassaroli$^*$}
\centerline { Physics Department, Northeastern University,  
Boston, MA 02115, USA}
\centerline {and}
\centerline{Physics Department and INFN, University of Perugia, Perugia, 
06100, Italy}
$$
$$
$$
$$
\centerline {\bf ABSTRACT}
$$ $$
A simple but general proof is presented to show that Lorentz 
covariance and 4-momentum conservation alone are sufficient 
to obtain muon oscillations in pion decay 
if the recoiling neutrinos oscillate.
$$ $$
$$ $$
$$ $$
$$ $$
$$ $$
$$ $$
$^*$ Present Address: Laboratory for Nuclear Science and Physics 
Department, Massachusetts Institute of Technology, Cambridge, MA 02139
\vfill
\eject
\centerline {\bf $\pi^-$ and $\mu^-$ Decays:}
\bigskip
    The purpose of this short note is to show that it is
impossible to find a $\pi^-$ momentum amplitude  
in the reaction $\pi^-\rightarrow \mu^-+\bar{\nu}_\mu $,
which yields a neutrino space-time oscillation without 
also yielding a space-time oscillation in the $\mu^-\rightarrow 
e^-+\bar{\nu}_e+\nu_\mu $ secondary vertex provided Lorentz 
invariance and four-momentum conservation are strictly maintained.
The proof goes as follows. Consider the double decay process 
$$
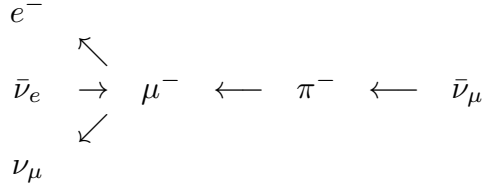
\matrix{e^- &\ &\ &\ &\ &\ &\ \cr
\ &\nwarrow &\ &\ &\ &\ &\ \cr
\bar{\nu}_e &\rightarrow &\mu^- &\longleftarrow &\pi^- &\longleftarrow 
&\bar{\nu}_\mu \cr
\ &\swarrow &\ &\ &\ &\ &\ \cr 
\nu_\mu &\ &\ &\ &\ &\ &\ \cr }
$$
in the  $\pi^-$ rest frame. Energy and momentum conditions read
$$
{\bf P}_{total}|\pi^->={\bf 0},\ \ H_{total}|\pi^->=M_\pi c^2|\pi^->.
$$
Thus for the outgoing space-time wave function (discrete quantum 
numbers implicit) defined as 
$$
\Psi ({\bf r}_\mu,t_\mu;{\bf r}_\nu,t_\nu)=
<{\bf r}_\mu,t_\mu;{\bf r}_\nu,t_\nu |\pi^->,
$$
energy and momentum conservation give 
$$
-i\hbar({\bf \nabla}_\mu +{\bf \nabla}_\nu)
\Psi ({\bf r}_\mu,t_\mu;{\bf r}_\nu,t_\nu)={\bf 0},
$$
and
$$
i\hbar\{(\partial/\partial t_\mu ) +(\partial/\partial t_\nu )\}
\Psi ({\bf r}_\mu,t_\mu;{\bf r}_\nu,t_\nu)=
M_\pi c^2 \Psi ({\bf r}_\mu,t_\mu;{\bf r}_\nu,t_\nu).
$$
The general solution with conserved energy and momentum is given by 
$$
\Psi ({\bf r}_\mu,t_\mu;{\bf r}_\nu,t_\nu)=
exp\{-i(M_\pi c^2/2\hbar )(t_\mu +t_\nu) \}
\psi ({\bf r}_\mu-{\bf r}_\nu,t_\mu-t_\nu).
$$
If $\psi ({\bf r}_\mu-{\bf r}_\nu,t_\mu-t_\nu)$ oscillates in 
$({\bf r}_\nu,t_\nu )$, then it also oscillates in 
$({\bf r}_\mu,t_\mu )$.
\bigskip 
\bigskip
\par \noindent
Widom@neuhep.physics.neu.edu 
\par \noindent
Yogi@neuhep.physics.neu.edu 

\bye